\documentclass[prl,twocolumn,floatfix,groupedaddress]{revtex4}
\usepackage{graphicx}
\usepackage{amsmath}
\usepackage[latin9]{inputenc}
\usepackage[T1]{fontenc}
\usepackage[english]{babel}

\newcommand{\be}{\begin{equation}}
\newcommand{\ee}{\end{equation}}
\newcommand{\ben}{\begin{eqnarray}}
\newcommand{\een}{\end{eqnarray}}

\newcommand{\nd}{\noindent}
\begin{document}

\title{Space-time correlations in urban population flows}

\author{A. Hernando$^1$, A. Plastino$^{2,\,3}$}

\affiliation{
$^1$ Laboratoire Collisions, Agr\'egats, R\'eactivit\'e, IRSAMC, Universit\'e Paul Sabatier 118 Route de Narbonne 31062 - Toulouse CEDEX 09, France\\
$^2$ National University La Plata, Physics Institute (IFLP-CCT-CONICET) C.C. 727, 1900 La Plata, Argentina\\
$^3$ Universitat de les Illes Balears and IFISC-CSIC, 07122 Palma
de Mallorca, Spain}


\begin{abstract}
 Evidences are presented concerning tantalizing regularities in cities'
population-flows in what regards to space and time correlations. 
The former exhibit a
distance-behavior (for large distances) compatible with the
inverse square law, following an overall Lorentzian dependence
with an scale-parameter of $74\pm6$~km. 
The later decay exponentially with a characteristic time of $17.2\pm1.3$ years. These features can be explained by a
dynamical model for cities' population-growth of a Lagevinian nature.
Numerical simulations based on the model confirm its applicability. The model also allows for the
identification of collective normal modes of  city-growth dynamics that can be empirically identified.
\end{abstract}
\maketitle

\nd The application of mathematical models to social
sciences has a long and distinguished history~\cite{1sta,1stb}. A
large number of studies show that the population-evolution in
urban agglomerations exhibits patterns that can be modeled by
mathematical laws (Refs.~\onlinecite{ciudad,ciudad2,thermoflow,empirical,gmodel,mob1} and references therein). In particular,
the interaction between cities (as measured by, for instance, the
number of crossed phone calls\cite{gmodel} or  human
mobility\cite{mob1}) displays  predictable characteristics. Thus,
it is plausible to conjecture that some kind of universality
underlies collective human behavior\cite{thermoflow,mob2}. 
The observation and detection of regular space-time patterns in urban-population evolution
may be viewed as constituting an important step towards understanding
collective, human dynamics. Indeed, the parametrization of such regularities could lead 
to a potential improvement of the present population-projection tools\cite{tools}.

\nd Based on official Census-data\cite{ine}, we analyzed the
time-evolution of the population of the $n=8116$ Spanish
Municipalities in a time-window of $13$ years, from 1998 to 2010,
with a total population $N$ of 47021031 in 2010.
We write the total population at year $t$ (setting $t=1$ for
the year 1998) as $N(t)=\sum_{i=1}^n  \,X_i(t)$, where $X_i(t)$ is the population of the $i$-th
Municipality at that year. In order to guarantee that we take into account
internal-flow effects we work with  relative-populations $x_i(t)=X_i(t)/N(t)$.
 The annual relative population change then reads \be
\dot{x}_i(t) = x_i(t+1)-x_i(t), \ee thus obtaining $T=12$ data
sets for this variable. We specifically focus attention upon  the pairs of data,
$\{x_i(t),\dot{x}_i(t)\}$ so as  to assess: 

\begin{itemize}
\item[I)] The mean value of the
population and the variance of its annual change in our time
window. As shown in Ref.~\onlinecite{thermoflow}, some dependence
of the later on the former is expected.
\item[II)] The spatial dependence
of the {\it Pearson product-moment correlation
coefficient}\cite{correc} for the annual change of each pair of
municipalities $i,j$.
\item[III)] The time-dependence of that correlation
coefficient for each pair of years of available data.
\item[IV)] Finally, we advance a Langevin equation\cite{langevin} for the evolution of the populations
 able to reproduce all these three characteristics.
\end{itemize}
Following the above scheme {\bf one has in step I)} the mean value of the population, the
mean annual change, and the variance of each population $i$ written as, respectively
\be
\begin{array}{c}
\displaystyle \langle x_i\rangle = \frac{1}{T+1}\sum_{t=1}^{T+1}x_i(t), ~~\langle\dot{x}_i\rangle = \frac{1}{T}\sum_{t=1}^{T}\dot{x}_i(t),\\
\displaystyle V[\dot{x}_i]=\langle[\dot{x}_i-\langle\dot{x}_i\rangle]^2\rangle =
\frac{1}{T}\sum_{t=1}^{T}(\dot{x}_i(t)-\langle\dot{x}_i\rangle)^2.
\end{array}
\ee
In the wake of  Ref.~\onlinecite{thermoflow} we plot in Fig.~1 the
pairs $\{\langle x_i\rangle,V[\dot{x}_i]/\langle x_i\rangle\}$.
The results nicely fit an expression of the type 
\be\label{eqV}
V[\dot{x}_i]/\langle x_i\rangle=\sigma^2\langle x_i\rangle +
\sigma_{1/2}^2 
\ee 
i.e., {\it proportional growth} ($\sigma$-term)
plus {\it finite-size noise} ($\sigma_{1/2}-$noise). A fit of the
data to that expression yields $\sigma=0.0119$ years$^{-1}$ and
$\sigma_{1/2}=6.9\times10^{-5}$ years$^{-1}$. Proportional growth becomes dominant for
large-population cities, while  finite-size noise becomes
dominant for low-populations. As shown in the  above cited
reference, for a network-based model proportional growth depends
on the actual structure of the social network, while  numerical
noise is a consequence of the Central Limit Theorem. It is then
to be expected that the former will convey information about the nature
of the system, while the later would constitute  just uncorrelated
noise.

\begin{figure}
\includegraphics[width=0.3\textwidth,trim = 30 30 170 400,clip]{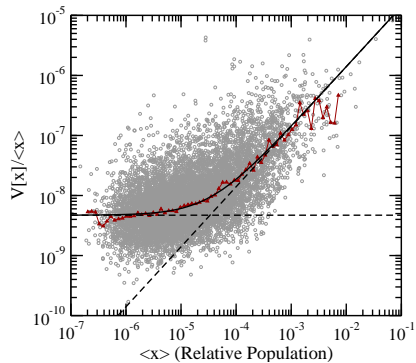}
\caption[]{$\{\langle x_i\rangle,V[x_i]/\langle x_i\rangle\}$ pairs for each Spanish municipality.
Red Triangles: median values. Solid black: fit to the median following Eq.~(\ref{eqV}). Dashed black lines: Finite-size's fluctuations
are constant, while the multiplicative regime is given by a straight line.
}
\end{figure}

\nd {\bf Let us pass now to step II), i.e., to spatial correlations.} The distance between cities is $r_{ij}$ for the $i,j$-th ones. The
correlation coefficient reads
 \be\label{cij}
c_{ij}=\mathrm{Cor}[\dot{x}_i,\dot{x}_j]=\frac{\mathrm{Cov}[\dot{x}_i,\dot{x}_j]}
{\sqrt{V[\dot{x}_i]V[\dot{x}_j]}}, \ee where we use the covariance 
$\mathrm{Cov}[\dot{x}_i,\dot{x}_j]=\langle[\dot{x}_i-\langle\dot{x}_i\rangle][\dot{x}_j-\langle\dot{x}_j\rangle]\rangle$. 
Fig.~2 displays the
empirical distribution $p(r,c)$ of the pairs $\{r_{ij},c_{ij}\}$ for
Spanish cities of population $>20000$ inhabitants ($x>4\times10^{-4}$, $n=396$), where
 proportional growth clearly dominates over  finite-size noise (see Fig.~1). 
We base bottom panel of Fig.~2 on an histogram of the
distance-correlation pairs, normalized along the $c$ direction
(using intervals of $\Delta \ln(r)=0.1$ and $\Delta c=1/15$) in a
window such that $1.7<\ln(r_{ij})<6$. We have applied a ``moving average'' of 10 points in the $\ln(r)$ direction
 so as to get a smoother surface for guiding the eye.
A clear dependence on the distance becomes evident. The ensuing distribution 
perfectly adjusts the expected correlation coefficient's distribution for a bivariate normal
distribution given in Ref.~\onlinecite{pc}. Using a finite number of data-points $T$,
we name this distribution as $P(c,C,T)$ where $c$ stands for the correlation-value that one might numerically
obtains using Eq. (\ref{cij}), and $C$ for the actual correlation value.
We assume for the latter the analytical form
\be\label{cr} 
C(r) = \frac{C(0)}{1+|r/r_0|^\alpha}, 
\ee
where $C(0)$, $r_0$, and $\alpha$ are adjustable parameters.
 $P(c,C,T)$ and $p(r,c)$ become then related via $p(r,c)=P[c,C(r),T]$ (see Fig.~2). We have found
that the empirical median value of the correlation, measured in the same intervals
$\Delta \ln(r)$ for a window $5<r_{ij}<1000$~km, nicely fits  the
above expression with $C(0)=0.254 \pm 0.009$, $r_0 = 74 \pm 6$ km,
 $\alpha = 2.1 \pm 0.3$, and a goodness coefficient of
$R^2=0.996$ (Fig.~2). Remarkably enough, for large distances one
has $C(r) \sim 1/r^2$, in agreement with the Gravity Model\cite{gmodel}. 
For the smallest cities ($x<2\times10^{-5}$) no evidence of spatial
correlation is encountered because of numerical noise $\propto
\sqrt{x}$, confirming our expectations. For mid-populated cities
we find a mixture between the two regimes. 
\begin{figure}
\includegraphics[width=0.3\textwidth,trim = 30 30 170 400,clip]{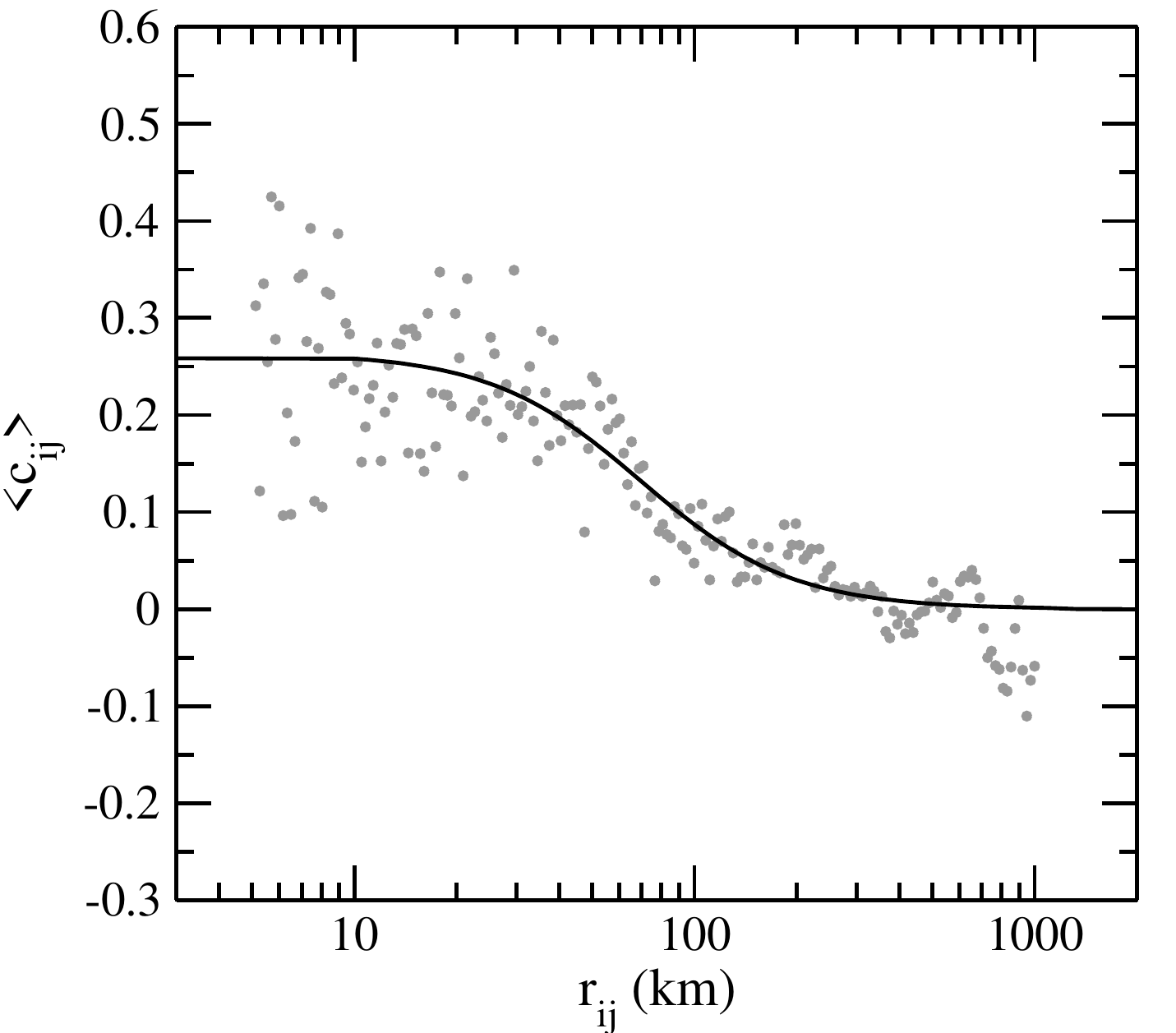}\\
\includegraphics[width=0.5\textwidth]{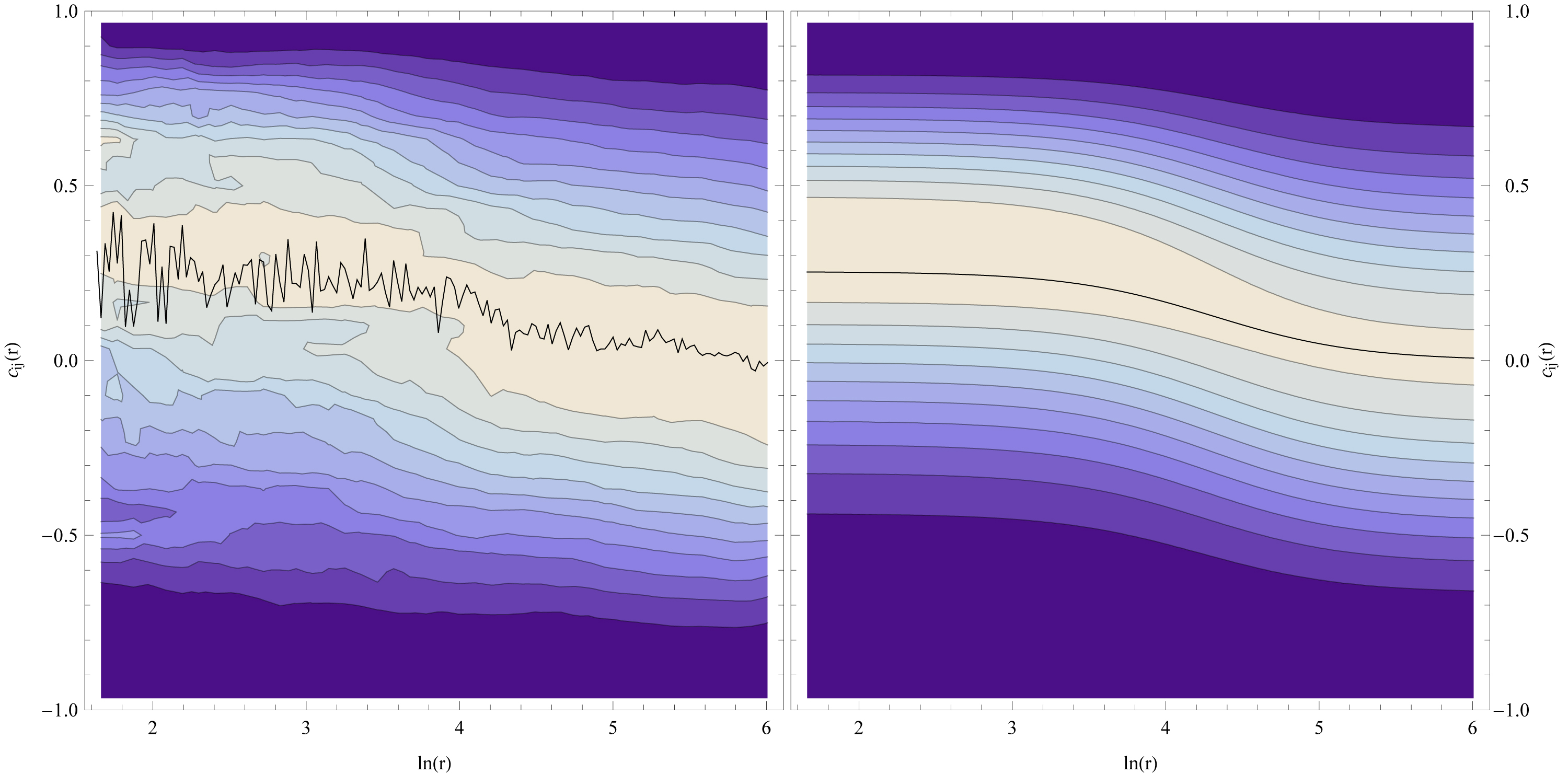}
\caption[]{Top panel: 
Empirical mean correlation vs. distance (points), fitted to Eq. (\ref{cr}) (line). Bottom panels: empirical $p(r,c)$ and theoretical
$P[c,C(r),T]$ distributions,  together with the pertinent mean value (black line).
}
\end{figure}

\nd {\bf We pass now to step III), time-correlations}. Consider i) the
$n$-cities-average and variance such that 
\be\label{sumt}
 \langle \dot{x}(t)\rangle =
\frac{1}{n}\sum_{i=1}^n \dot{x}_i(t),~
V[\dot{x}(t)]
=
\frac{1}{n}\sum_{i=1}^n[\dot{x}_i(t)-\langle\dot{x}(t)\rangle]^2,
\ee
and ii) the ensuing correlation coefficient. 
We estimate time-correlations via the average
\be\label{cdt} c(\Delta t)=\frac{1}{T-\Delta
t}\sum_{t=1}^{T-\Delta t}\mathrm{Cor}[\dot{x}(t),\dot{x}(t+\Delta t)]. 
\ee
Here $\Delta t$
adopts the discrete values 1, 2, 3,$\ldots$, T-1. Fig.~3 depicts
results for the major Spanish cities, as in the previous case.
The ensuing temporal dependence can be  parameterized using the
expression $c(\Delta t)=ae^{-\gamma\Delta t}$. We find
$1/\gamma=17.2\pm1.3$~years and $a=0.70\pm0.02$, with a  goodness
coefficient $R^2=0.997$. Again, correlations are not clearly
discernible in the case of small-population cities, telling us that a
finite-size noise, proportional to $\sqrt{x}$, is indeed
independent of time. The transition between both regimes is 
depicted in Fig.~3. We display the empirical value of
$c(\Delta t=1)$ as a function of the population in intervals of
$\Delta \ln(x)=0.25$. Growth from $0$ to $\sim0.7$ is clearly
visible.
\begin{figure}[t!]
\includegraphics[width=0.5\textwidth,trim = 65 560 20 50,clip]{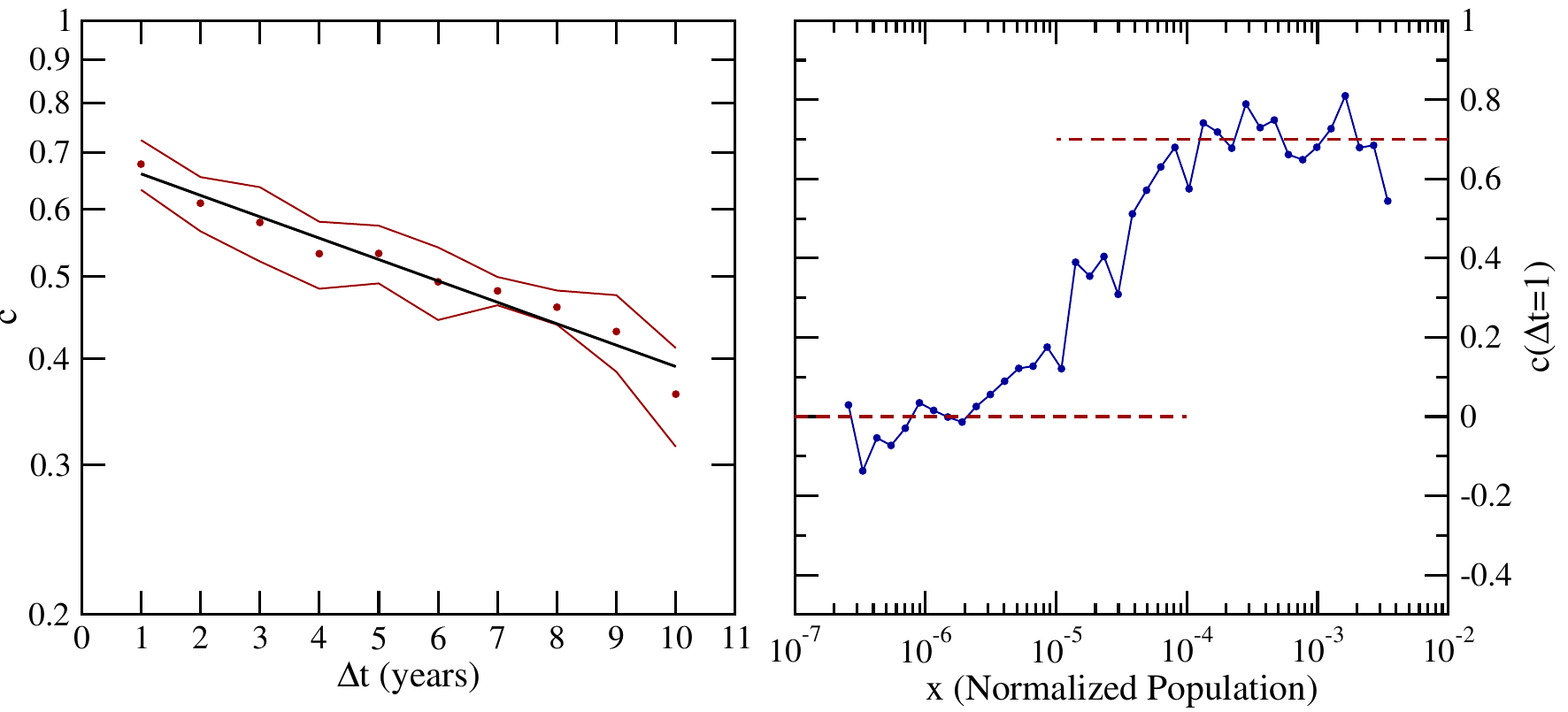}
\caption[]{Left: Empirical correlation vs. time (dots) fitted to an exponential function (straight line).
The red lines represent the standard deviation of the sum of Eq.~(\ref{sumt}). Note the
log-scale on the vertical axis. Right: One-year correlation
vs. relative population, where the transition from uncorrelated finite-size noise regime (zero correlation)
to correlated proportional growth regime ($C(1)\sim0.7$) is clearly appreciated.
}
\end{figure}

\nd {\bf In our final (IV) step}, we advance a dynamical model able to
reproduce the three empirical results we have encountered above,
namely,  i) uncorrelated numerical noise, independent from $r,t$,
for small cities together with proportional growth for large
populations, ii) Lorentz-type spatial correlations, and iii)
exponential time-correlation.  We propose here the Langevin-like
equation \cite{langevin} \ben
\dot{x}_i(t) &=& x_i(t)v_i(t)+\sqrt{x_i(t)}w_i(t)\\
\dot{w}_i(t) &=& W_i(t)\\
\dot{v}_i(t) &=& F_i(t) - \gamma_i v_i(t)\\
F_i(t) &=& \sum_{i'} R_{ii'}f_{i'}(t), \een where i) $W_i(t)$
characterizes a Wiener-process ($\langle w_i(t) w_j(t+\Delta
t)\rangle=V[w]\delta_{ij}\delta(\Delta t)$, where $V[w]$ is the corresponding variance), ii) $\gamma_i$ are
dumping-parameters to be empirically determined, with regards to
time-correlations, iii) the $R_{ii'}$ are matrix-elements related
to spatial correlations, and iv) $f_{i'}(t)$ are stochastic
forces. We assume 
that the forces fluctuate and are
independent of each other, being of the form \be
\mathrm{Cov}[f_i(t),f_j(t+\Delta t)] = V_f\delta_{ij}\delta(\Delta t),
\ee where  $V_f$ stands for the variance of $f$. For the total
force $F$ the covariance $\mathrm{Cov}[F_i(t),F_j(t+\Delta t)]\equiv K $    is \be  K= V_f\sum_{i'}R_{ii'}R_{ji'}\delta(\Delta t) =
V_fQ_{ij}\delta(\Delta t). \ee The matrix
$Q=R\cdot R^T$ is normalized in such a fashion that its diagonal
elements are all equal to unity, and thus $\mathrm{cor}[F_i(t),F_j(t)]=Q_{ij}$. 
 Since $w_i(t)$ and $f_i(t)$ are not correlated by definition, 
Eq.~(\ref{eqV}) in automatically fulfilled from the variance of $\dot{x}_i(t)$
with $\sigma^2=V[v_i]$ and $\sigma_{1/2}^2=V[w]$.
 If the $w-$term is the dominant one (low population) there is
no correlation among cities (nor in time) since we deal with a
Wiener process (no memory, either). For large populations the
$v-$term dominates and we have $C_{ij}(\Delta
t)=\mathrm{Cor}[\dot{x}_i(t),\dot{x}_j(t+\Delta
t)]=\mathrm{Cor}[v_i(t),v_j(t+\Delta t)]$ with a 
general solution to the Langevin equation for $v$  
\be
v_i(t) = e^{-\gamma_it}v_i(0) + \int_0^td\tau e^{-\gamma_i(t-\tau)}F_i(\tau).
\ee
Since the initial time is arbitrary, we assume $t\rightarrow\infty$
so as to obtain the $v-$correlation \be\label{esso} C_{ij}(\Delta
t)=\mathrm{cor}[v_i(t),v_j(t+\Delta t)] = Q_{ij}e^{-\gamma_j\Delta
t}\frac{2\sqrt{\gamma_i\gamma_j}}{\gamma_i+\gamma_j}. \ee Note
that $2\sqrt{\gamma_i\gamma_j}/(\gamma_i+\gamma_j) \in [0,1]$,
and its mean value depends on the distribution of the $\gamma-$values.
\begin{figure}[t!]
\includegraphics[height=0.3\textwidth,trim = 20 30 120 350,clip]{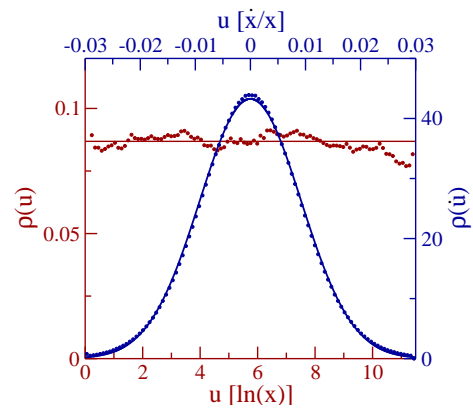}
\caption[]{
Equilibrium density distribution
of the log-population and the relative growth obtained in  the simulation (dots), compared
with the thermodynamical prediction Eq. (\ref{rho}) (lines).}
\end{figure}
\nd {\it Our goal would now be nicely achieved} if we could show that
$Q_{ij}=1/(1+|r_{ij}/r_0|^2)$. To attain such result, let us start
with a result of Ref.~\onlinecite{gmodel}. 
The number of phone-calls between two cities can be fitted to our
Lorentzian shape. If we assume such a pattern for the information
flow in a human social network, we may regard  the forces $F$ as
resulting from  the average of the stochastic forces $f$ weighted
by that Lorentzian function. Thus, $F$ becomes a ``coarse-grained"
force. This is reflected by the definition \be R_{ij} =
\frac{2[2/\pi r_0^2]^{1/4}}{1+4|r_{ij}/r_0|^2}. \ee Let us
consider for our derivation of $Q$ the continuous limit
$x_i\rightarrow x(\mathbf{r})$, with $\mathbf{r}$ a planar spatial
coordinate. $x(\mathbf{r})$ represents the relative population at
$\mathbf{r}$, and the total normalized population becomes
$1=\int_Sd\mathbf{r}x(\mathbf{r}),$ where  $S$ is the pertinent
region's area. Since we deal now with the coordinates $\mathbf{r}$
and $\mathbf{r}'$ instead of the indexes $i,j$, the $R$
matrix-elements are a function $R(|\mathbf{r}-\mathbf{r}'|)$ and
the total coarse-grained force follows the convolution ($\otimes$)
\be 
F(\mathbf{r},t) = R(\mathbf{r})\otimes f(\mathbf{r},t) =
\int_Sd\mathbf{r}'\frac{2[2/\pi r_0^2]^{1/4}f(\mathbf{r}',t)}{1+4|(\mathbf{r}-\mathbf{r}')/r_0|^2}.
\ee Since the convolution of two Lorentzians of equal scale is
also a Lorentzian with twice that scale-parameter, we find for the
forces-correlation \be
\begin{array}{r}\displaystyle \mathrm{cor}[F(\mathbf{r}),F(\mathbf{r}')]=
R(|\mathbf{r}-\mathbf{r}'|)\otimes R(|\mathbf{r}-\mathbf{r}'|)\\
\displaystyle=Q(|\mathbf{r}-\mathbf{r}'|)=\frac{1}{1+|(\mathbf{r}-\mathbf{r}')/r_0|^2},
\end{array}
\ee i.e., {\it the result we wished to reach}. {\bf  We can now interpret our equations in the following fashion}:
i) initially, some stochastic, independent forces $f_i$ act on the
cities' populations, ii) the information-flow within the
population can be characterized via a Lorentzian distribution, so
that the effective total force becomes the convolution of the
$f_i$ with the distribution, iii) the observed correlations are
thus Lorentzian, and  at large distances decay as the square of
the distance.  Finally, setting  $\gamma_i=1/17~\forall i$ and $r_0=74$~km, our model can
reproduce the empirical correlations. The result $C(0)<1$ can be
attributed to both i) some numerical, uncorrelated noise and ii) to the
empirical distribution of the dumping parameters $\gamma_i$ [as
can be verified by looking at  Eq.~(\ref{esso})]. 

\nd{\bf As an application} we have performed a suitable simulation. We
randomly select  $n=100$ positions for population-centers on a square surface of
side  $L=250$~km 
(see Supplementary Figure). We took $V_f=10^{-5}$~years$^{-2}$, $\gamma=1/17$~years$^{-1}$,
and $r_0=74$~km, forcing the population to evolve within the range
$X_0<X<X_M$, with $X_0=1$ and $X_M=10^4$~people. We do not include finite-size
effects $W$ for simplicity ($V[w]=0$).
The computational cost of solving our Langevin equation can be reduced via
 a normal-mode treatment: define a change-of-basis matrix  $A$ such that
$R$ (and $Q$) become diagonal. This generates new variables
$u'_i(t)=\sum_{i'}A_{ii'}\log[X_i(t)]$ whose motion-equations become
\be
\dot{u}'_i(t) = v'_i(t);~~
\dot{v}'_i(t) = \sqrt{\varepsilon_i}f'_i(t) - \gamma v'_i(t),
\ee with  $v'_i(t)=\sum_{i'}A_{ii'}v_i(t)$,
$f'_i(t)=\sum_{i'}A_{ii'}f_i'(t)$, and $\sqrt{\varepsilon_i}$ is
the $i$-th eigenvalue of $R$ (with  $\varepsilon_i$ that of $Q$).
One easily checks that the forces $f'$ are statistically
equivalent to those indicated by $f$ [i.e., 
$\langle f'_i(t)f'_j(t+\Delta t)\rangle = V_f\delta_{ij}\delta(\Delta t)$], so that the simulation involves directly
the random generation of $f'$, without having to actually effect the
basis-change. The variables $u'_i(t)$ evolve in independent
fashion, representing normal-mode evolution. The presence of
$\sqrt{\varepsilon_i}$ accounts for different mode-equilibrations
between  $f'$ and the dumping $\gamma$, which might be
conceived as originating mass-factors. The Supplementary Figure displays the first
four modes in such a way that the color at the Municipality  $i$
represents the coefficient $A_{ii'}$ for the eigenvector $i'$
(we also shown the same decomposition for Catalonia (Spain), using the empirical value of $r_0$ found above).
Our simulations use Verlet-integration\cite{verlet}
with an interval $\delta t=1$.  Equilibrium states are detected,
compatible with the thermodynamics developed in Ref.~\onlinecite{thermoflow}.
 We verified that in equilibrium
$\langle [v(t)]^2\rangle=V_f/2\gamma=8.5\times10^{-5}$ and that the
normalized equilibrium distribution  $\rho(X,\dot{X})dXd\dot{X}$ follows the
tenets applicable to for a thermal system confined in a constant volume. As a
matter of fact, defining the log-population $u=\log(X)$ ($\dot{X}/X=\dot{u}$) one has\cite{thermoflow}
\be\label{rho}
\rho(X,\dot{X})dXd\dot{X} =
\rho(u,\dot{u})dud\dot{u} =
\frac{\sqrt{\beta/2\pi}}{\log{X_0/X_M}}e^{-\frac{\beta\dot{u}}{2}},
\ee
where $\beta=2\gamma/V_f$, as depicted in Fig.~4. 

{\bf Summing up}, our model can be fine-tuned to any
socio-geographical area via the dumping parameters $\gamma_i$ and
the inclusion, in the forces, of any empirically-known factor that
may affect population growth. There is also room to include in
the correlation matrix $Q$ any other empirically-known
distance-independent correlation. The  model may  improve on 
the  predictive mathematical tools available today \cite{tools}.
Also, the study of the past evolution of the population in terms
of  normal models ould lead to a deeper understanding of 
collective human behavior at the macro-scale.

\begin{figure*}[t!]
\includegraphics[width=\textwidth]{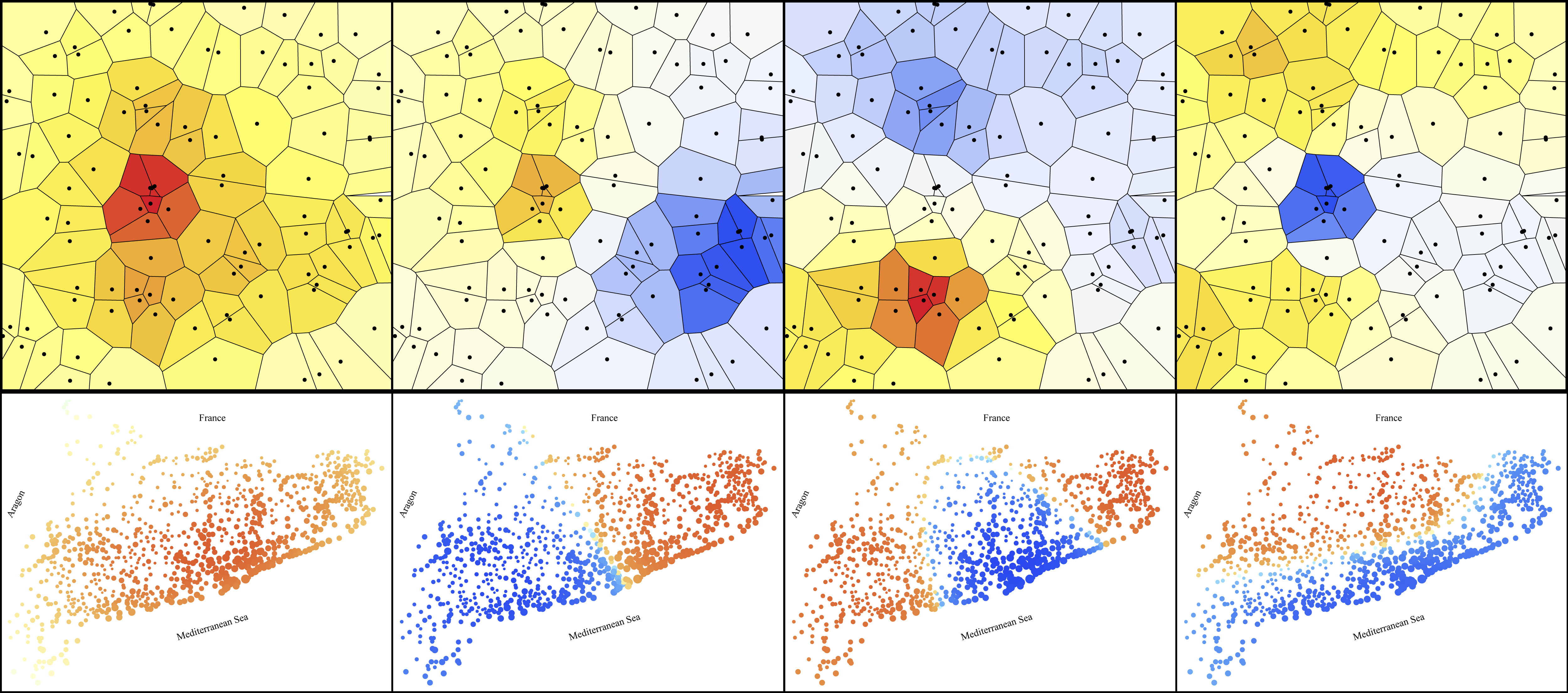}

{\bf Supplementary Figure.} Top: Components of the first four eigenvectos of the simulated system (from white to red: positive values; from white to blue: negative values). The surface of each municipality is defined by its Voronoi area (http://en.wikipedia.org/wiki/Voronoi\_diagram).
Bottom: Components of the first four eigenvectos for Catalonia, using the modelled spatial correlations (see text). The radii of the circles are proportional to the log-population.
\end{figure*}

\end{document}